\documentclass[aps,preprintnumbers,nofootinbib,superscriptaddress,floatfix]{revtex4}

\usepackage{amssymb}
\usepackage{amsmath}
\usepackage{bm}
\usepackage{slashed}
\usepackage{datetime}
\usepackage{mciteplus}
\usepackage{multirow}
\usepackage{color}
\usepackage[dvipsnames]{xcolor}
\usepackage[utf8]{inputenc}
\usepackage[normalem]{ulem}
\usepackage{graphicx}
\usepackage{hyperref}
\newcommand{\bT}{\bm{b }_{T}}
\newcommand{\qT}{\bm{q}_{T}}
\newcommand{\kperp}{\boldsymbol{k}_\perp}
\pdfoutput=1 
             
\begin{document}
\allowdisplaybreaks[2]

\title{New insights from the flavor dependence of \\ 
quark transverse momentum distributions in the pion \\ \vspace{0.2cm}
\normalsize{\textmd{The \textbf{MAP} (Multi-dimensional Analysis of Partonic distributions) Collaboration}}}

\author{Lorenzo Rossi}
\thanks{Electronic address: lorenzo.rossi3@unimi.it --\href{https://orcid.org/0000-0002-8326-3118}{ORCID: 0000-0002-8326-3118}}
\affiliation{Dipartimento di Fisica, Universit\`a di Milano, Via Celoria 16, 20133 Milan, Italy} 
\affiliation{INFN, Sezione di Milano, Via Celoria 16, 20133 Milan, Italy}

\author{Alessandro Bacchetta}
\thanks{Electronic address: alessandro.bacchetta@unipv.it --\href{https://orcid.org/0000-0002-8824-8355}{ORCID: 0000-0002-8824-8355}}
\affiliation{Dipartimento di Fisica, Universit\`a di Pavia, Via Bassi 6, 27100 Pavia, Italy} 
\affiliation{INFN, Sezione di Pavia, Via Bassi 6, 27100 Pavia, Italy}

\author{Matteo Cerutti}
\thanks{Electronic address: matteo.cerutti@cea.fr --\href{https://orcid.org/0000-0001-7238-5657}{ORCID: 0000-0001-7238-5657}}
\affiliation{IRFU, CEA, Université Paris-Saclay, F-91191 Gif-sur-Yvette, France}

\author{Marco Radici}
\thanks{Electronic address: marco.radici@pv.infn.it --\href{https://orcid.org/0000-0002-4542-9797}{ORCID: 0000-0002-4542-9797}}
\affiliation{INFN, Sezione di Pavia, Via Bassi 6, 27100 Pavia, Italy}

\begin{abstract}
We update our previous extraction of transverse momentum distributions of unpolarized quarks in the pion by implementing a more comprehensive description of theoretical uncertainties and, for the first time, by exploring possible differences among quark flavors. 
We extract such distributions from all available data for unpolarized pion-nucleus Drell-Yan processes, where the cross section is differential in the transverse momentum of the final lepton pair. The cross section involves transverse momentum distributions in the nucleon, that we consistently take from our previous studies.
{ We find indications that inside the pion the valence quark may have wider tail at large transverse momenta with respect to the sea quark, a result that could not be exposed by the simple flavor-independent analysis. }
\end{abstract}


\maketitle
\section{Introduction}

The pion and the proton are the simplest bound states that can be formed with the quark and gluon (collectively, parton) degrees of freedom of Quantum ChromoDynamics (QCD). The former is a boson, the latter a fermion. More importantly, the pion is the Goldstone boson of the spontaneous breaking of chiral symmetry, see e.g.~\cite{Horn:2016rip}. Hence, it is crucial to understand how its internal structure generates such differences with the bound state of the proton. However, due to the limited availability of high-energy scattering data, the structure of the pion is still much less understood than that of the proton.

Since the 1990s, several extractions from experimental data of parton distribution functions (PDFs) in the pion have been released~\cite{Gluck:1991ey,Sutton:1991ay,Gluck:1997ww,Gluck:1999xe, Wijesooriya:2005ir,Aicher:2010cb,Barry:2018ort,
Novikov:2020snp,Cao:2021aci,Barry:2021osv,Bourrely:2022mjf,Barry:2023qqh,Pasquini:2023aaf,Good:2025nny,Kotz:2023pbu,Kotz:2025lio,Schweitzer:2010tt}. But PDFs describe parton  distributions along the hadron’s longitudinal momentum. As such, they provide only a one-dimensional information. A full three-dimensional picture can be achieved through transverse-momentum dependent PDFs (TMD PDFs), which include information also on the parton intrinsic transverse momentum. Such information for the pion is available in the literature mostly through model calculations~\cite{Pasquini:2014ppa,Noguera:2015iia,Lorce:2016ugb,Bacchetta:2017vzh,Ceccopieri:2018nop,Ahmady:2019yvo,Kaur:2019jfa,Shi:2020pqe,Cheng:2024gyv}. Only few extractions of pion TMD PDFs are  available~\cite{Wang:2017zym,Vladimirov:2019bfa,Cerutti:2022lmb,Barry:2023qqh} that are limited to data only for the Drell-Yan (DY) process. 

In this paper, we update our previous work of Ref.~\cite{Cerutti:2022lmb} in two important ways: 
i) we perform a more comprehensive analysis of the uncertainty in the results by propagating the theoretical error from the full Monte Carlo set of input PDFs; ii) for the first time, we investigate the differences among transverse-momentum distributions of different quark flavors inside the pion. We analyze all available data for the DY lepton-pair production in $\pi^-$-nucleus collisions. For proton–proton collisions, DY data are known to have only weak sensitivity to flavor differences between the annihilating quark–antiquark pairs. However, for pion-proton collisions the situation is different because antiquarks can be valence quarks in the pion. The cross section differential in the lepton-pair transverse momentum involves the convolution of an unpolarized quark TMD PDF in $\pi^-$ and of an unpolarized quark TMD PDF in the proton. 
For the latter, we use the recent result of the MAP collaboration~\cite{Bacchetta:2024qre} where the flavor dependence of proton TMD PDFs is also considered. We consistently use the same universal Collins--Soper evolution kernel for both pion and proton TMD PDFs.

\section{Formalism}
\label{sec:formalism} 

We consider the DY process $h_A(P_A) + h_B(P_B) \to \ell^+ + \ell^- + X$ where a lepton pair with total four-momentum $q$ and invariant mass $Q=\sqrt{q^2}$ is inclusively produced from the collision of the hadrons $h_A$ and $h_B$ with four-momenta $P_A$ and $P_B$, respectively, and center-of-mass squared energy $s=(P_A+P_B)^2$.

If the mass $M$ of the hadrons and the transverse component $\qT$ of the lepton pair momentum with respect to the collision axis satisfy the conditions $M^2 \ll Q^2$ and $\qT^2 \ll Q^2$, the differential cross section can be written as
\begin{equation}
\label{e:DYxs}
\frac{d\sigma^{DY}}{d|\qT|dydQ} 
= \mathcal{P}\, \frac{8\pi\alpha_{em}^2}{9Q^3} |\qT| x_A x_B \mathcal{H}^{DY}(Q;\mu) \sum_a c_a(Q^2) \int d|\bT| |\bT| J_0(|\qT||\bT|) \hat{f}_1^a(x_A, \bT^2; \mu, \zeta_A) \hat{f}_1^{\bar{a}}(x_B, \bT^2; \mu, \zeta_B) \, ,
\end{equation}
where $\mathcal{P}$ is a phase space factor accounting for potential cuts on the final leptons kinematics, $\alpha_{em}$ is the electromagnetic coupling constant, $y$ is the lepton pair rapidity, $x_{A,B} = e^{\pm y} \, Q/\sqrt{s}$ are the longitudinal momentum fractions carried by annihilating partons $a$ and $\bar{a}$,  $\mathcal{H}^{DY}(Q;\mu)$ is the perturbative hard factor accounting for the virtual part of the scattering and depending on $Q$ and the renormalization scale $\mu$, and finally the summation runs over all active quark flavors with $c_a(Q^2)$ their electroweak charges~\cite{Bacchetta:2022awv,Cerutti:2022lmb}. { We emphasize that, in contrast to collinear factorization, gluons do not contribute through TMD PDFs at leading power in the cross section. Their contribution enters only through the collinear PDFs appearing in the matching convolutions (see Eq.~\eqref{e:TMD_matching})}.

In Eq.~\eqref{e:DYxs}, $\hat{f}_1^a$ represents the Fourier transform of the TMD PDF for the unpolarized quark $a$ inside the hadron $h_A$. It depends on the quark longitudinal momentum fraction $x_A$, on the variable $\bT$ Fourier-conjugated to the quark transverse momentum, on the renormalization scale $\mu$ and the rapidity scale $\zeta_A$~\cite{Collins:2011zzd} (and, similarly, for the annihilating partner $\bar{a}$ inside hadron $h_B$ with momentum $x_B$ at rapidity scale $\zeta_B$). Such dependence is controlled by a set of coupled differential evolution equations~\cite{Bacchetta:2022awv}. A convenient choice for the initial scale is $\mu_i = \sqrt{\zeta_i} = \mu_b(|\bT|) = 2 e^{-\gamma_E}/|\bT|$ (where $\gamma_E$ is the Euler constant) because it avoids the insurgence of large logarithms, particularly in the rapidity evolution kernel~\cite{Collins:2011zzd}. In order to prevent $\mu_b$ from exceeding $Q$ at small $|\bT|$ or from approaching the QCD Landau pole $\Lambda_{\text{QCD}}$ at large $|\bT|$, we adopt the same {\it ad hoc} prescription of Refs.~\cite{Bacchetta:2017gcc,Bacchetta:2019sam,Bacchetta:2022awv,Cerutti:2022lmb,Bacchetta:2024qre,Bacchetta:2025ara,Bacchetta:2024yzl} replacing $\mu_b$ with $\mu_{b_*} = 2 e^{-\gamma_E}/b_*(\bT)$ where
\begin{equation}
b_*(\bT) = b_{\text{max}} \, \left( \frac{1-e^{-|\bT|^4/b_{\text{max}}^4}}{1-e^{-|\bT|^4/b_{\text{min}}^4}} \right)^{1/4} \, , 
\label{e:bstar}
\end{equation}
with
\begin{equation}
b_{\text{max}} = 2 e^{-\gamma_E} \, \mbox{GeV}^{-1}\, , \quad b_{\text{min}} = 2 e^{-\gamma_E} / \mu \, .
\end{equation}
Then, the TMD PDF can be reformulated as
\begin{equation}
\label{e:TMDb*}
\hat{f}_{1}^a(x,\bT^2;\mu,\zeta) = 
\left[\frac{\hat{f}_{1}^a(x,\bT^2;\mu,\zeta)}{\hat{f}_{1}^a(x,b_*(\bT^2);\mu,\zeta)}\right] 
\hat{f}_{1}^a(x,b_*(\bT^2);\mu,\zeta) 
\equiv f_{\text{NP}}^a(x,\bT^2;\zeta) \, \hat{f}_{1}^a (x,b_*(\bT^2);\mu,\zeta) \, ,
\end{equation}  
namely it can be split in a nonperturbative component $f_{\text{NP}}$ and a perturbatively calculable part. 

Using the Operator Product Expansion at the input scale $\mu_{b_*}$, the perturbative component can be matched onto its corresponding collinear PDF $f_1$:
\begin{equation}
\label{e:TMD_matching}
\hat{f}_1^a(x,b_*(\bT^2);\mu_{b_*},\mu_{b_*}^2) = \sum_b \int_x^1 \frac{dx'}{x'} C^{ab}(x',b_*(\bT^2);\mu_{b_*},\mu_{b_*}^2) \, f_1^b\bigg(\frac{x}{x'};\mu_{b_*}\bigg) 
\equiv [C \otimes f_1]^a (x,b_*(\bT^2);\mu_{b_*},\mu_{b_*}^2) \, ,
\end{equation}  
where the matching coefficients $C$ can be  perturbatively expanded in powers of the strong coupling constant $\alpha_s$. The final expression for the perturbative component of the TMD PDF in Eq.~\eqref{e:TMDb*} results 
\begin{equation}
\label{e:pert_TMD}
\hat{f}_1^a(x,b_*(\bT^2);\mu,\zeta) =  
[C \otimes f_1]^a (x,b_*(\bT^2);\mu_{b_*},\mu_{b_*}^2) \,  
\exp\bigg\{ \int_{\mu_{b_*}}^{\mu} \frac{d\mu'}{\mu'}\, \gamma\big(\mu',\zeta\big) \bigg\} \,  
\bigg(\frac{\zeta}{\mu_{b_*}^2}\bigg)^{K(b_*(\bT^2), \, \mu_{b_*})/2} \, ,
\end{equation} 
where 
\begin{equation}
\label{e:gamma_mu}
\gamma(\mu,\zeta) = \gamma_F\big(\alpha_s(\mu)\big) - \gamma_K\big(\alpha_s(\mu)\big)\, \ln\frac{\sqrt{\zeta}}{\mu} \, ,
\end{equation}  
and $\gamma_K$ is the cusp anomalous dimension, $K$ is the perturbative part of the Collins--Soper kernel (accounting for the anomalous dimension associated with the renormalization of rapidity divergences~\cite{Collins:2011zzd}), and $\gamma_F\big(\alpha_s(\mu)\big) = \gamma\big(\mu,\mu^2\big)$~\cite{Bacchetta:2019sam}.

The nonperturbative component of the pion TMD PDF in Eq.~\eqref{e:TMDb*} is parametrized as in Ref.~\cite{Cerutti:2022lmb}:
\begin{equation}
\label{e:pif1NP}
f_{\text{NP}}^a (x,\bT^2; \zeta) = 
e^{-g_{\pi}^a (x) \, \bT^2/4} \bigg{[} \frac{\zeta}{Q_0} \bigg{]}^{-g_2^2 \, \bT^2/4} \, ,
\end{equation}  
where the $x$-dependent Gaussian width $g_{\pi}^a$ is parametrized as
\begin{equation}
\label{e:Gausswidth}
g_{\pi}^a (x) = N_{\pi}^a \,\frac{x^{\sigma_{\pi}^a} \, (1-x)^{(\alpha^a_{\pi})^2}}{\hat{x}^{\sigma_{\pi}^a} \, (1-\hat{x})^{(\alpha^a_{\pi})^2}} \; ,
\end{equation}  
with $\hat{x} = 0.1$. Apart from the three free parameters $N_\pi^a,\, \sigma_\pi^a,\,\alpha_\pi^a$ of the Gaussian function, the last part of Eq.~\eqref{e:pif1NP} represents the nonperturbative part of the Collins--Soper kernel and involves two more parameters, $Q_0$ and $g_2$. The $Q_0$ is an arbitrary scale and $g_2$ drives the nonperturbative evolution at large $\bT$ in the rapidity scale. Because of the universality of the Collins-Soper kernel, both $Q_0$ and $g_2$ are consistently taken from our previous extraction of proton TMD PDFs in Ref.~\cite{Bacchetta:2024qre}. 

Finally, the TMD PDF depends on the input of the corresponding collinear PDF (through its perturbative part, see Eq.~\eqref{e:pert_TMD}). Consistently with our previous work~\cite{Cerutti:2022lmb}, we use the xFitter~\cite{Novikov:2020snp} collinear PDF set. However, in order to efficiently propagate the PDF uncertainties onto the final observable~\cite{Bacchetta:2024qre} we have converted the xFitter Hessian set of PDFs into a Monte Carlo set of 100 PDF replicas using the procedure outlined in Refs.~\cite{Watt:2012tq,Hou:2016sho}, and employing the public {\tt mcgen} code~\cite{mcgen}~\footnote{ We verified that the Monte Carlo set of 100 PDF replicas is sufficient to reproduce the average and variance of the original xFitter Hessian set of PDFs.}. {
Although several pion PDF extractions are available and exhibit some level of tension among them, we do not expect this to affect our analysis, since they are mutually compatible in the kinematic region covered by the dataset considered here. A more systematic study of the impact of different collinear PDF sets is left for future work.}

\section{Data}

\begin{figure}[h]
\centering
\includegraphics[width=0.48\textwidth]{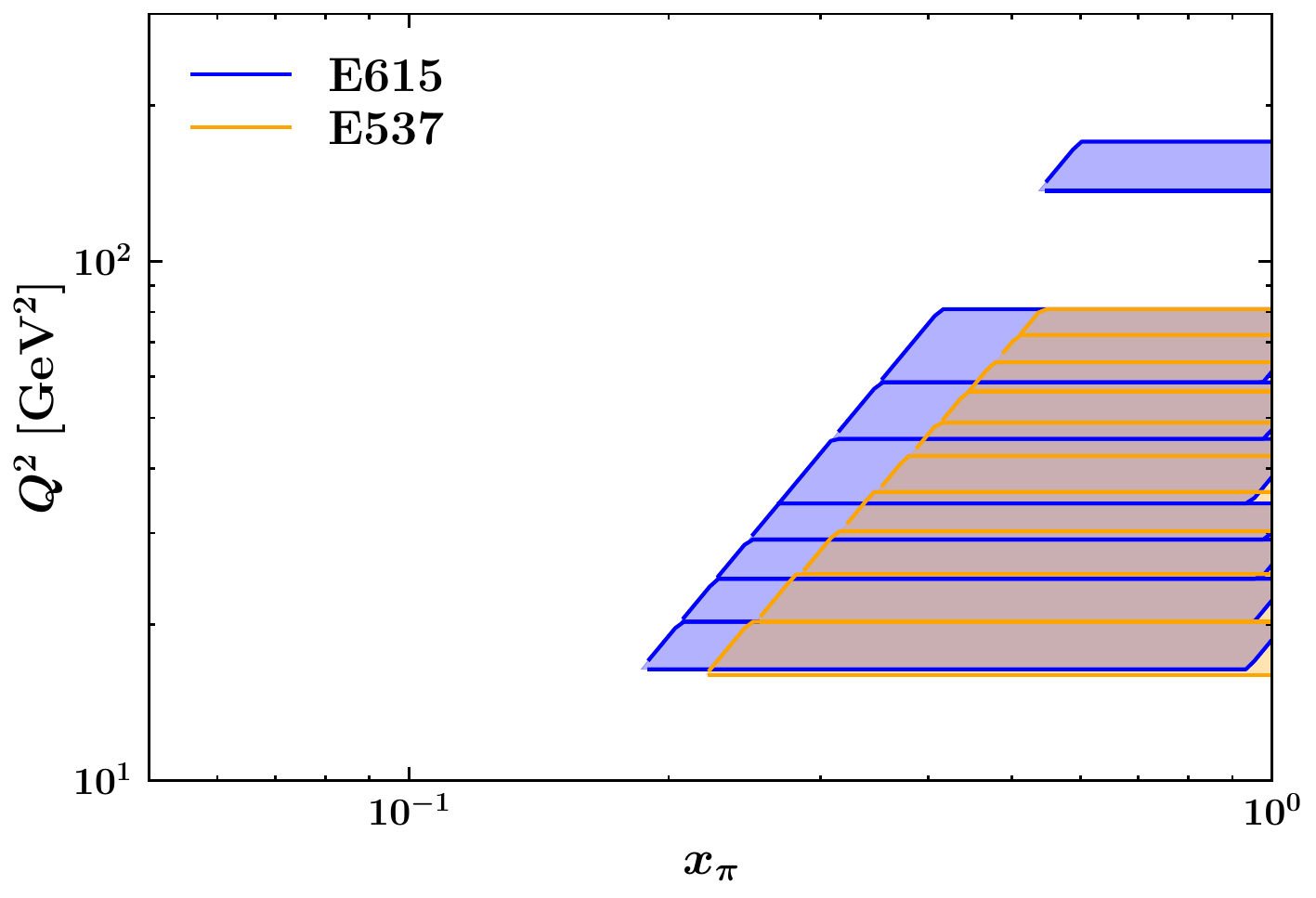} \hspace{0.3cm}
\includegraphics[width=0.48\textwidth]{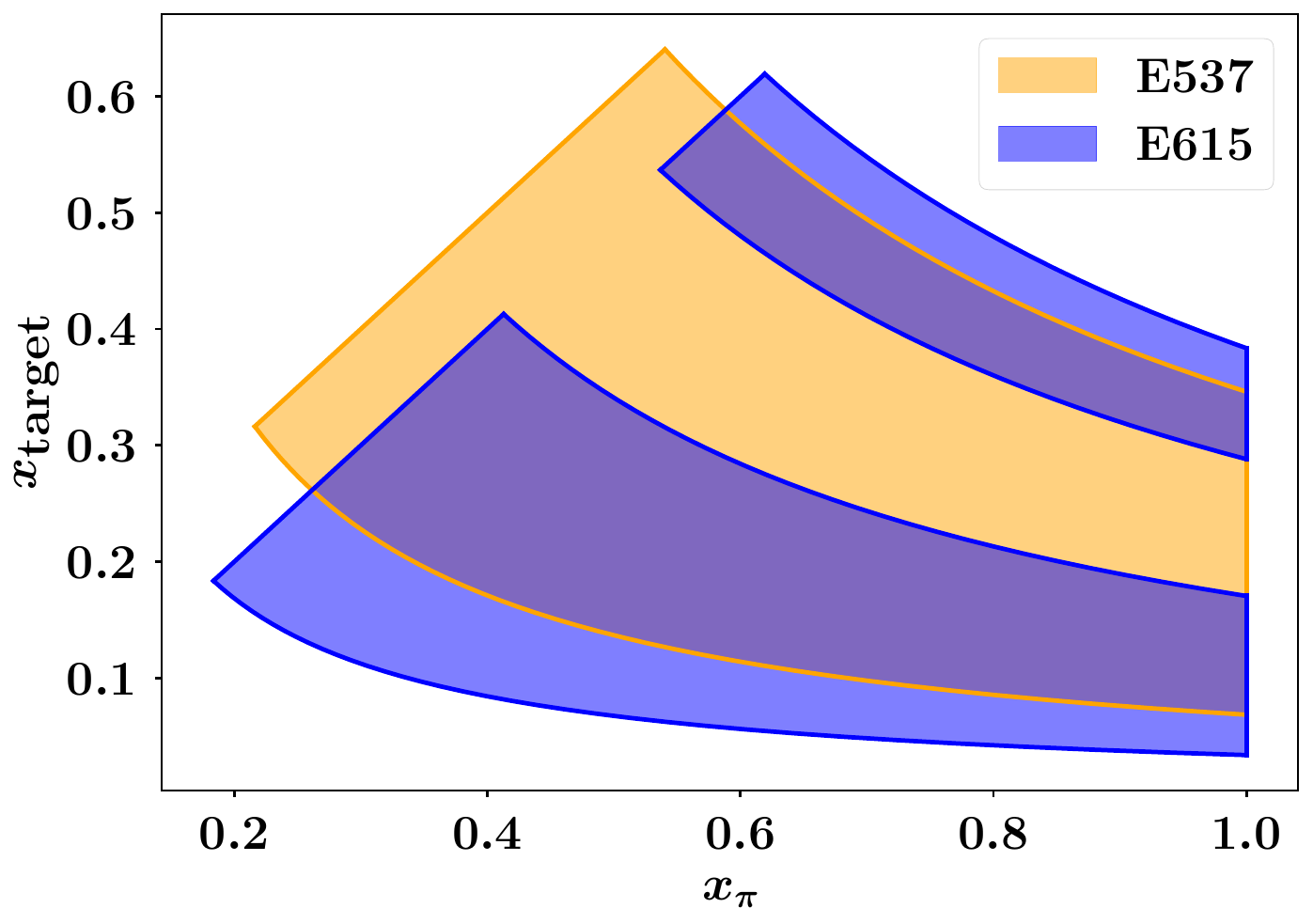}
\caption{Kinematic coverage of experimental data used in this analysis from the E615~\cite{Conway:1989fs} and E537~\cite{Anassontzis:1987hk} Drell-Yan experiments with $\pi^-$ beams. Left panel for the $(x_\pi, Q^2)$ plane, right panel for the $(x_\pi, x_{\text{target}})$ plane.}
\label{f:kin_coverage_pion}
\end{figure}

In this analysis, we use experimental data from the two experiments E615~\cite{Conway:1989fs} and E537~\cite{Anassontzis:1987hk} where DY lepton pairs were measured from the collision of $\pi^-$ beams and tungsten fixed targets (for these experiments, $\mathcal{P}=1$ in Eq.~\eqref{e:DYxs}). The kinematic coverage of these data sets is illustrated in Fig.~\ref{f:kin_coverage_pion}, where in the left and right panels  
the data are represented in the $(x_\pi, Q^2)$ and $(x_\pi, x_{\text{target}})$ planes, respectively.  Fig.~\ref{f:kin_coverage_pion} shows that data explore a  kinematic region limited to quark fractional momenta $x_\pi$ in the valence region. 

The cross section of Eq.~\eqref{e:DYxs} is based on the constraint $\qT^2 \ll Q^2$ to grant the applicability of TMD factorization. To fulfill this constraint, we apply to experimental data the following kinematic cut 
\begin{equation}
	\label{e:cut_pion_new}
	\frac{|\qT|}{Q} < 0.2 + \frac{0.6~\text{GeV}}{Q} \; .
\end{equation}
This choice is slightly more stringent than in our previous work~\cite{Cerutti:2022lmb}, is closer to the analogous cut adopted in proton TMD extractions, and ensures the applicability of TMD factorization without excessively reducing the number of included data points.

In addition, we exclude in the E615 data set the bins for $9.00 \leq Q \leq 11.70$ GeV that cover the region of the $\Upsilon$ resonance as made in the previous fit. In Tab.~\ref{t:pion_data}, we list the data before ($N_{\text{dat}}$) and after the cuts ($N_{\text{cut}}$) together with measured observables and kinematic ranges for each experiment. 
We deliberately consider the cross section differential in $Q$ (rather than in $x_F$) because 
having a well-defined $Q$ bin for each data point ensures that the kinematical cut of Eq.~\eqref{e:cut_pion_new} can be evaluated consistently. Using data binned only in $x_F$ would instead mix events spanning a broad range in $Q$, effectively producing a single inclusive $Q$ bin and making the application of the cut considerably more ambiguous.

\begin{table}[h]
	\begin{center}
		\begin{tabular}{|c||c|c|c|c|c|c|}
			\hline
			Experiment  & $N_{\mathrm{dat}}$  & 
            $N_{\mathrm{cut}}$  &
            Observable & $\sqrt{s}$ [GeV] & $Q$ range [GeV] & $x_F$ range  \\
			\hline \hline
			E615 & 
            155 & 
            41 & $d^2 \sigma / dQ d|\qT|$ &
                        21.8 & 4.05~-~13.05 & 0.0~-~1.0 \\ \hline
			E537 & 
            150 &
            60 & $d^2 \sigma / dQ d\qT^2$ & 15.3 & 4.0~-~9.0 & -0.1~-~1.0 \\
			\hline
		\end{tabular}
		\caption{Summary of experimental data  included in this analysis: for each experiment, the number of data points before ($N_{\mathrm{dat}}$) and after kinematic cuts ($N_{\mathrm{cut}}$), the measured observable, the center-of-mass energy $\sqrt{s}$, the range of invariant mass $Q$, and the integration range in $x_F = x_\pi - x_{\text{target}}$.}
		\label{t:pion_data}
	\end{center}
\end{table}

For both experiments, the measurements are affected by large statistical and systematic errors. 
These uncertainties are treated in this analysis as fully correlated. 
Following the same methodology used in previous works by the MAP Collaboration~\cite{Bacchetta:2022awv,Cerutti:2022lmb,Bacchetta:2024qre,Bacchetta:2024yzl,Bacchetta:2025ara},
the theoretical uncertainties inherited by the input PDF are also included 
 in the computation of the $\chi^2$.

\section{Results}
\label{s:Results}

In this section, we present the results of the phenomenological extraction of unpolarized quark TMD PDFs in the pion, which are obtained by fitting all the available DY data involving pions. Tables with grids of the obtained TMD PDFs will be made publicly available on the MAP Collaboration website.\footnote{\href{https://github.com/MapCollaboration}{https://github.com/MapCollaboration}}

Unlike the MAPTMD24 analysis~\cite{Bacchetta:2024qre}, we cannot reach full N$^3$LL accuracy because the collinear pion PDF sets are available only at NLO accuracy. Therefore, we extract the pion TMD PDFs at N$^3$LL$^-$ accuracy, following a terminology similar to that used in the MAPTMDPion22 analysis~\cite{Cerutti:2022lmb}.

As done in previous works by the MAP Collaboration~\cite{Cerutti:2022lmb,Bacchetta:2022awv,Bacchetta:2024qre,Bacchetta:2024yzl,Bacchetta:2025ara}, the error analysis is carried out using the bootstrap method, where an ensemble of 100 Monte Carlo replicas of experimental data is fitted. Specifically, for the $i$-th replica we utilize the $i$-th element from the Monte Carlo set of the NNPDF and xFitter collinear PDFs for proton and pion, respectively; in this way, we propagate the uncertainty of the PDFs into the final error estimate. To ensure consistency with MAPTMD24, the parameters from the $i$-th replica of MAPTMD24 is used and kept fixed in the proton TMD PDF. For the same reason, we have not considered 
nuclear modifications, although they have been recently studied in other extractions~\cite{Alrashed:2021csd,Alrashed:2023xsv}. They are anyway very small at the momentum fractions $x$ covered by E615 and E537 data. 

In the bootstrap method, the complete statistical information is contained within the entire set of replicas of the extracted TMD PDFs. However, we select the $\chi^2$ value of the best fit to the unfluctuated data, $\chi^2_0$, as the indicator of the quality of our fit (we refer to this as the $\chi^2$ of the central replica). 
As in our previous works, we further decompose $\chi^2_0$ as
$\chi^2_0 = \chi^2_D + \chi^2_\lambda$, where $\chi^2_D$ is the contribution of uncorrelated uncertainties and the penalty $\chi^2_\lambda$ is related to correlated uncertainties  (for the explicit expressions of $\chi^2_D$ and $\chi^2_\lambda$, see   Refs.~\cite{Cerutti:2022lmb,Bacchetta:2022awv}). 

\begin{table}
\centering
\begin{tabular}{|c||c|c|c|c|}
\hline
\textbf{Experiments} & $N_{\text{cut}}$ & $\chi_D^2 / N_{\text{cut}}$  &
$\chi_\lambda^2 / N_{\text{cut}}$       & $\chi_0^2 / N_{\text{cut}}$
\\ \hline \hline
E537             & 41 &  0.88      & 0.04 & 0.92 \\ \hline
E615           & 60 &  1.34     & 0.18 & 1.52 \\ \hline
Total & 101 &1.16   &  0.12 & 1.28 \\ \hline
\end{tabular}
\caption{Breakup of the $\chi^2_0$ of the central replica  for the MAPTMDPion25 FI fit, normalized to the number of data points $N_{\text{cut}}$ surviving the kinematic cuts, into the uncorrelated ($\chi^2_D$) and correlated ($\chi^2_\lambda$) uncertainties, for both E537 and E615 experiments.}
\label{t:chi2_pion_FI}
\end{table}

\subsection{Flavor-independent approach}

In this section, we describe a fit that we conventionally name MAPTMDPion25 Flavor Independent (FI). In this case, we use the same nonperturbative model of Eq.~\eqref{e:pif1NP} for all flavors, and we conventionally use the down quark PDF in Eq.~\eqref{e:TMD_matching} as the valence distribution in $\pi^-$. In Tab.~\ref{t:chi2_pion_FI} we show the quality of the fit. The $\chi_0^2$, normalized to the number of data points surviving the kinematical cuts, is split into its components for both E537 and E615 experiments. The quality of the fit of E615 data is systematically lower than for E537, probably because of the smaller uncorrelated uncertainties of the former with respect to the latter. The total result is 1.28. The major contribution to this value primarily arises from the uncorrelated component $\chi_D^2 = 1.16$. Although the $\chi_0^2$ value is relatively higher than in fits for proton TMDs, it is lower than our previous extraction~\cite{Cerutti:2022lmb} and Ref.~\cite{Vladimirov:2019bfa}, and it is comparable with the result of Ref.~\cite{Barry:2023qqh} (in this latter work, however, the authors have refitted also the collinear PDFs). 

We do not observe any normalization issue, at variance with what reported in Refs.~\cite{Cerutti:2022lmb,Vladimirov:2019bfa}. While in Ref.~\cite{Piloneta:2024aac} the authors argue for the inclusion of kinematic power corrections to accurately describe low-energy DY data, in our analysis we do not need to include such ingredients, consistently with other independent studies~\cite{Barry:2023qqh,Aslan:2024nqg} and with our successful description of fixed-target DY data in global fits involving unpolarized proton TMDs~\cite{Bacchetta:2024qre,Bacchetta:2025ara}. 

\begin{table}[h]
    \centering
    \begin{tabular}{|c|c|c|c|c|c|} \hline
          Parameter & Value & Parameter & Value & Parameter & Value  \\ \hline \hline
       $N_{\pi}$ & 0.035 $\pm$ 0.021 &$\alpha_{\pi}$ & 1.83 $\pm$ 0.24 & $\sigma_{\pi}$ & 3.62 $\pm$ 0.82 \\ \hline
    \end{tabular}
    \caption{Mean values and uncertainties at 68\% C.L. of the free parameters of the MAPTMDPion25 FI fit. \label{t:param_pion_FI} }
\end{table}
In Tab.~\ref{t:param_pion_FI}, we list the mean values and related uncertainties at the 68\% confidence level (C.L.) for the three fitted parameters. The $\alpha_\pi$ parameter controls the behavior of the pion TMD PDF at large $x$ (see Eq.~\eqref{e:Gausswidth}), where basically all of the available fixed-target DY data are concentrated; it is not surprising, then, that it can be constrained with an error as small as $\sim 13$\%. Conversely, the parameter $N_\pi$ has a large uncertainty because it drives the TMD PDF at $x = 0.1$ (see, again, Eq.~\eqref{e:Gausswidth}), where there are no DY data involving pions. A more careful scrutiny is needed for the parameter $\sigma_\pi$. In principle, it is associated with the low-$x$ behavior of the pion TMD PDF, a region not constrained by the experimental data included in the fit. Its relatively small uncertainty ($\sim 23$\%) is connected to a non negligible correlation with the ``$\alpha$ parameter" of the TMD PDF of the proton. We believe that this does not represent an issue in our analysis, since it pertains to a kinematic range where currently there are no data. In this region, the extracted pion TMD PDF should rather be viewed as an extrapolation of the model and, therefore, interpreted with caution.

In summary, the errors in Tab.~\ref{t:param_pion_FI} are in general not comparable with the ones that can be obtained in the extraction of proton TMDs, but they can be reduced to reasonably small values in the kinematical regions where data can constrain our functional form. 

\begin{figure}[h]
\centering
    \includegraphics[width=1.0\linewidth]{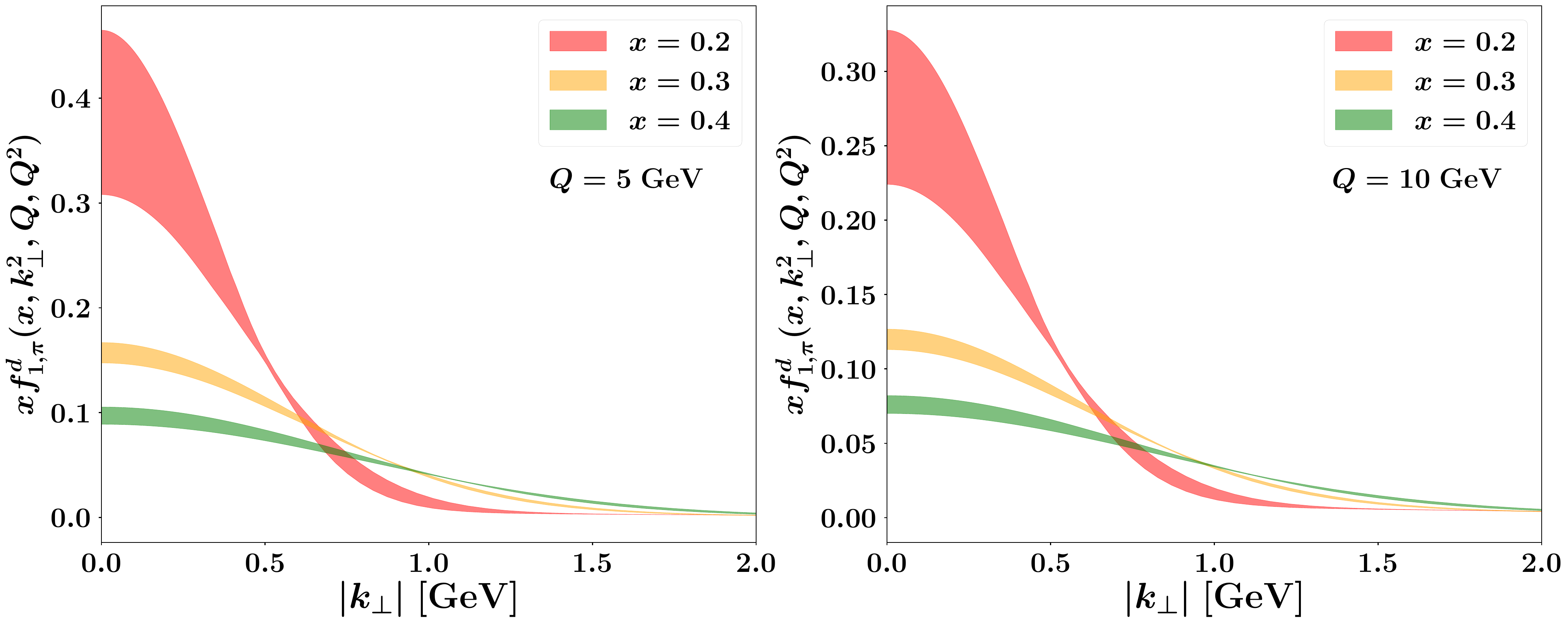}
    \caption{The TMD PDFs for an unpolarized down quark in $\pi^-$, extracted from the MAPTMDPion25 FI fit as functions of the partonic transverse momentum $ |\kperp| $ at $ x = 0.2$, $0.3$, and $ 0.4 $. Left panel for $ \mu = \sqrt{\zeta} = Q = 5$ GeV, right panel for $ 10$ GeV. The uncertainty bands corresponding to the central 68\% C.L.}
    \label{f:TMDPDF_pion_FI}
\end{figure}

In Fig.~\ref{f:TMDPDF_pion_FI}, we present the TMD PDF for an unpolarized down quark in $\pi^-$, as a function of the quark transverse momentum $|\kperp|$ for $x = 0.2$, $0.3$, and $0.4$, in order to cover the region of experimental data illustrated in Fig.~\ref{f:kin_coverage_pion}. The left panel shows the results at $\mu = \sqrt{\zeta} = Q = 5$ GeV, the right panel at $10$ GeV. The error bands correspond to the central 68\% C.L. band and include the propagated uncertainties of the collinear PDFs. 

At $x = 0.2$, the TMD PDF exhibits the largest uncertainties, particularly at low $|\kperp|$ values, because this kinematic region is only partially covered by the available data sets. Upcoming data from the COMPASS collaboration is anticipated to provide significant improvements in constraining pion TMD PDFs in this portion of the phase space.

\subsection{Flavor-dependent approach}

In this section, we present the first extraction of unpolarized quark TMD PDFs in the pion with flavor-dependent transverse momentum distributions. We adopt the same strategy of our previous extraction of proton TMDs~\cite{Bacchetta:2024qre}, namely we employ the same functional form of the flavor-independent case but we assign separate parameters for the different flavors. Given the limited sensitivity of data involving only the negatively charged pion $\pi^-$, we adopt one parametrization for the $d$ and $\bar{u}$ quarks, and another for all remaining sea quarks. 

\begin{table}[h]
    \centering
    \begin{tabular}{|c|c|c|c|c|c|} \hline
          Parameter & Value & Parameter & Value & Parameter & Value  \\ \hline \hline
       $N_{\pi}^d$ & 0.16$\pm$ 0.10 &$\alpha_{\pi}^d$ & 1.32 $\pm$ 0.42 & $\sigma_{\pi}^d$ & 2.11 $\pm$ 0.99 \\ \hline 
       $N_{\pi}^{{\rm sea}}$ & 62.5 $\pm$ 62 &$\alpha_{\pi}^{{\rm sea}}$ & 21.7 $\pm$ 20.7 & $\sigma_{\pi}^{{\rm sea}}$ & 121.4 $\pm$ 120.6 \\ \hline 
    \end{tabular}
    \caption{Mean value and uncertainties at 68\% C.L. of the free parameters of the flavor-dependent MAPTMDPion25 FD fit. \label{t:param_pion_FD}}
\end{table}

The fit, conventionally named MAPTMDPion25 Flavor Dependent (FD), involves six parameters, {\it i.e.} the three parameters $N_\pi^a, \, \sigma_\pi^a, \, \alpha_\pi^a$ for $a=d,\, {\rm sea}$. The obtained values are listed in Tab.~\ref{t:param_pion_FD}. The results for the valence channel are quite different from the flavor-independent case listed in Tab.~\ref{t:param_pion_FI} because they are influenced by the sea quarks. 
The latter channel is not well constrained by data, with errors at the level of almost 100\%. Nevertheless, their influence is very important because it allows one to have a more realistic picture of the valence components in the pion, including their larger uncertainties with respect to the results in Tab.~\ref{t:param_pion_FI}. We believe that this is an important outcome of this work. As a final remark, the larger uncertainties for the parameters of the valence channel in Tab.~\ref{t:param_pion_FD} call for new data as a necessary step towards better constraining the corresponding pion TMD PDF.

\begin{table}[h]
\centering
\begin{tabular}{|c||c|c|c|c|}
\hline
\textbf{Experiments} & $N_{\text{cut}}$ & $\chi_D^2 / N_{\text{cut}}$  &
$\chi_\lambda^2 / N_{\text{cut}}$       & $\chi_0^2 / N_{\text{cut}}$
\\ \hline \hline
E537             & 41 &  0.90      & 0.04 & 0.94 \\ \hline
E615           & 60 &  1.23     & 0.17 & 1.40 \\ \hline
Total & 101 &1.10   &  0.12 & 1.22 \\ \hline
\end{tabular}
\caption{Breakup of the $\chi^2_0$ for the central replica of the MAPTMDPion25 FD fit, normalized to the number of data points $N_{\text{cut}}$ surviving the kinematic cuts, into the uncorrelated ($\chi^2_D$) and correlated ($\chi^2_\lambda$) uncertainties, for both the E537 and E615 experiments.}
\label{t:chi2_pion_FD}
\end{table}

In Tab.~\ref{t:chi2_pion_FD}, we show the quality of the MAPTMDPion25 FD fit by listing the breakup of $\chi^2_0/N_{\text{cut}}$ into its uncorrelated ($\chi^2_D$) and correlated ($\chi^2_\lambda$) components, for both the E537 and E615 experiments.  Introducing flavor dependence in the analysis yields 
{ a total $\chi^2_0/N_{\text{cut}} = 1.22$, compared to the flavor-independent 1.28 in Tab.~\ref{t:chi2_pion_FI}. It is a small improvement, but it amounts to a $\Delta \chi^2 = 6.1$ in the total $\chi^2$ that corresponds, for three additional parameters, to a significance larger than 1~$\sigma$ (based on the likelihood-ratio test).}
This improvement is less pronounced than in the proton case discussed in a similar context~\cite{Bacchetta:2024qre}. The difference most likely stems from the DY dataset for pions being limited and weakly sensitive to flavor, a feature that emerges also in extractions of proton TMDs using only DY data~\cite{Moos:2023yfa}. 
However, we remark that there is a non-negligible improvement in the description of the E615 data (compare the 1.40 of Tab.~\ref{t:chi2_pion_FD} with the corresponding 1.52 of Tab.~\ref{t:chi2_pion_FI}), mostly related to a much better description of the lowest two bins in $Q$ (from $\chi^2_D/N_{\text{cut}} = 2.04$ to 1.26, and from 2.44 to 1.63, respectively).

\begin{figure}[h] 
\centering 
\includegraphics[width=0.85\linewidth]{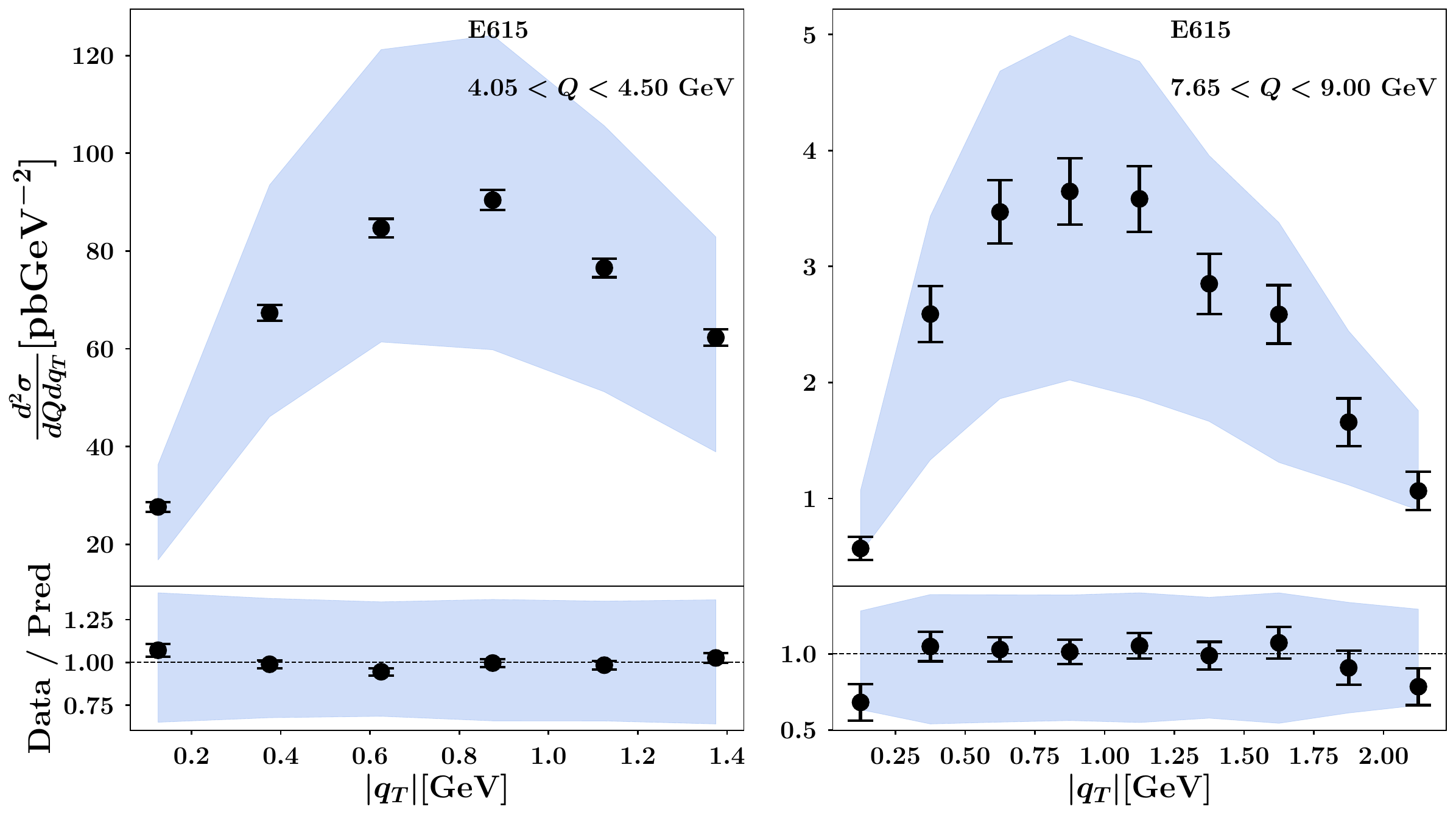} 
\caption{Upper panels for the differential Drell-Yan cross section as function of $|\qT|$ from the E615 experiment; left panel for the $4.05 < Q < 4.50$ GeV bin, right panel for the $7.65 < Q < 9$ GeV bin. Black dots for the experimental data points, colored band for our fit. Lower panels illustrate the ratio between theoretical predictions and experimental data. Uncertainty bands correspond to the central 68\% C.L.} 
\label{f:E615} 
\end{figure}

In Fig.~\ref{f:E615}, we present the comparison between the results of our fit (represented by the colored bands) and the experimental data (depicted by black points) for the $4.05 < Q < 4.5$ GeV (left panel) and $7.65 < Q < 9$ GeV (right panel) bins from the E615 dataset. The upper panels display the differential DY cross section as a function of the virtual boson's transverse momentum $|\qT|$, while the lower panels show the ratio between theoretical predictions and experimental data. The uncertainty bands, corresponding to the central 68\% C.L., demonstrate that our fit accurately reproduces the shape of the experimental data. The broadness of the error bands primarily reflects the large correlated systematic error in the normalization of the data points, which amounts to $\sim 16$\% (this error is not explicitly shown in the plots).

\begin{figure}[h] 
\centering 
\includegraphics[width=0.85\linewidth]{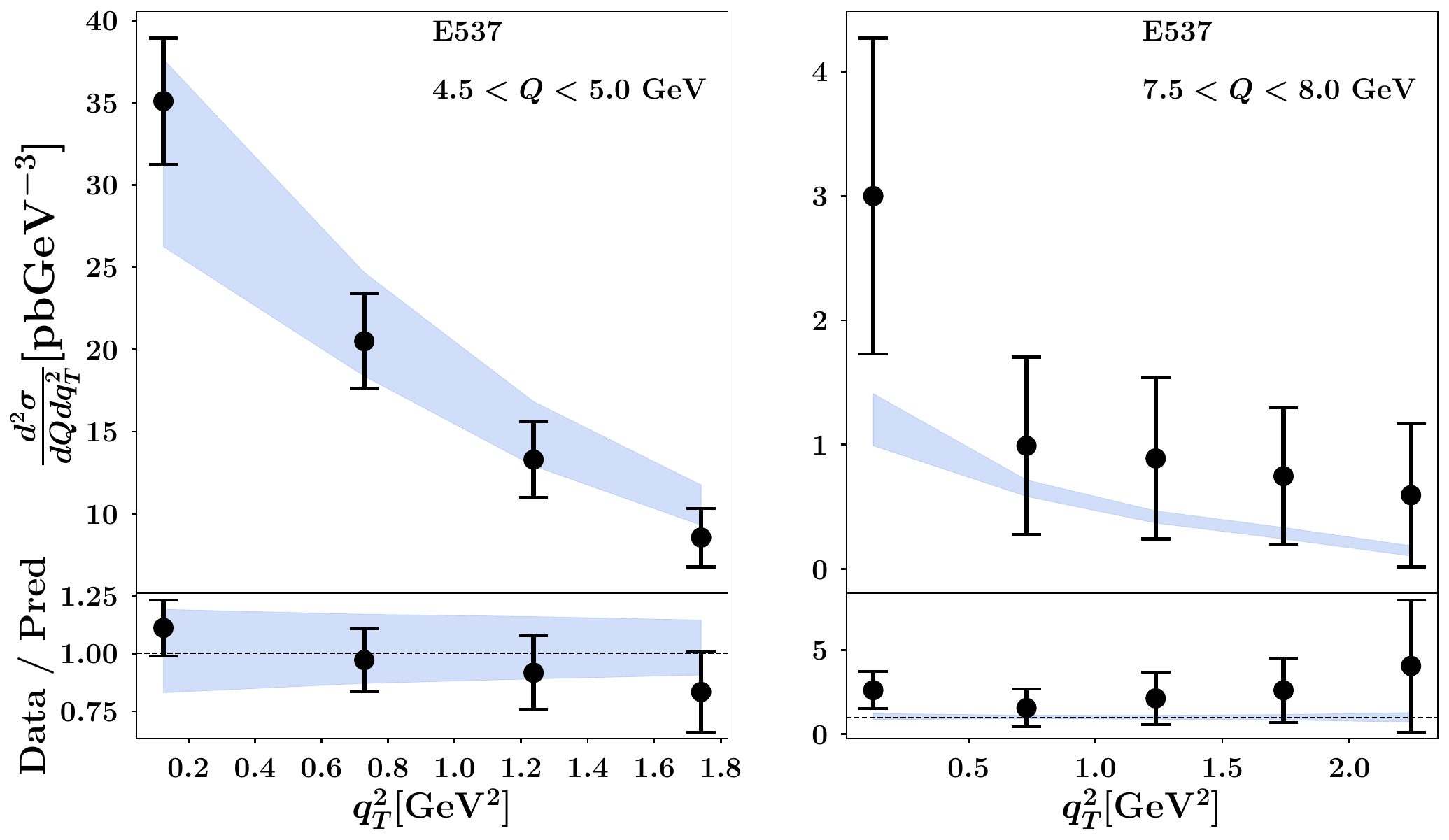} 
\caption{Same as in previous figure but for the $4.5 < Q < 5$ GeV and $7.5 < Q < 8$ GeV bins from the E537 experiment.} 
\label{f:E537} 
\end{figure}

In Fig.~\ref{f:E537}, we make the same comparison with the same notation and conventions as in Fig.~\ref{f:E615} but for the $4.5 < Q < 5$ GeV and $7.5 < Q < 8$ GeV bins from the E537 experiment. The DY cross section is now plotted as a function of $\qT^2$ rather than $|\qT|$. The central 68\% C.L. uncertainty bands are narrower than in the previous figure because the E537 data is affected by a smaller systematic error, which is reflected in a much smaller penalty $\chi^2_\lambda / N_{\text{cut}}$ in Tab.~\ref{t:chi2_pion_FD}. 

\begin{figure}[h]
    \centering
    \includegraphics[width=0.99\linewidth]{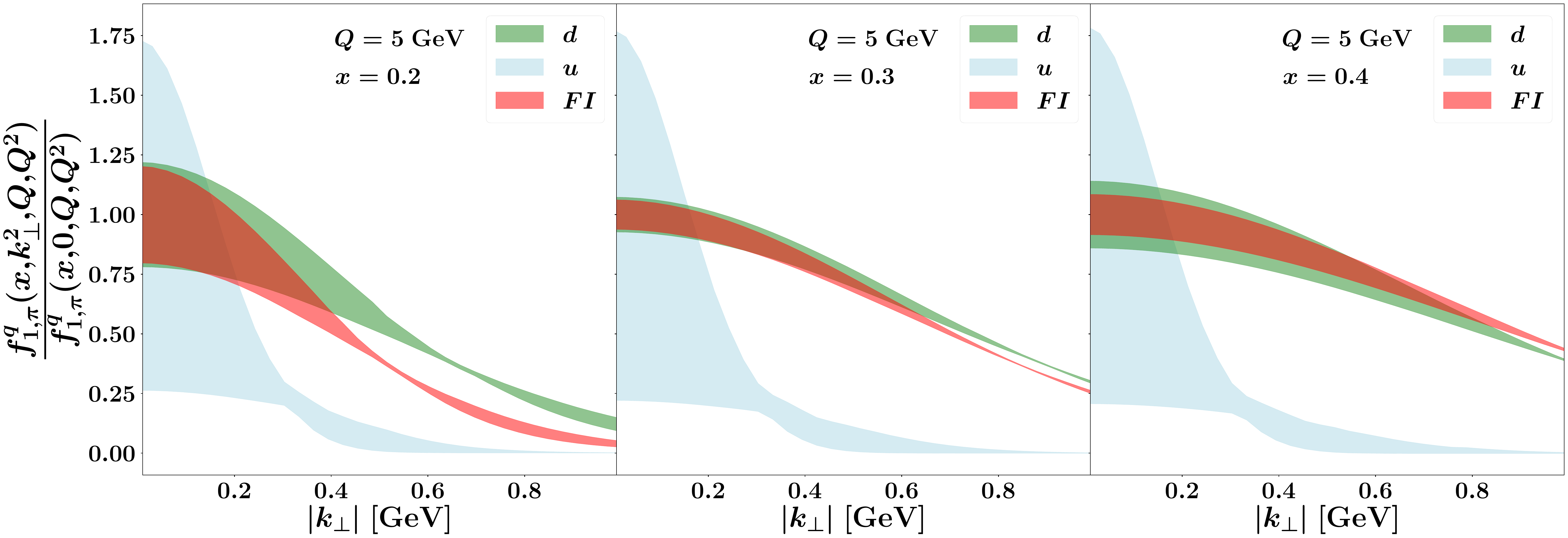}
    \caption{The TMD PDFs in $\pi^-$ as functions of $|\boldsymbol{k}_\perp|$, normalized at $|\boldsymbol{k}_\perp| = 0$ at $\mu = \sqrt{\zeta} = Q = 5$ GeV for $x = 0.2$ (left panel), $x = 0.3$ (central panel), and $x = 0.4$ (right panel). Red band from the flavor-independent MAPTMDPion25 FI fit, light-blue and green bands for the $u$ and $d$ quarks, respectively, from the flavor-dependent MAPTMDPion25 FD fit. All uncertainty bands correspond to the central 68\% C.L.}
    \label{f:TMDPDF_pion_FD_norm}
\end{figure}

In Fig.~\ref{f:TMDPDF_pion_FD_norm}, we show the TMD PDFs in $\pi^-$ resulting from the flavor-dependent MAPTMDPion25 FD fit for down and up quarks, as function of $|\boldsymbol{k}_\perp|$ at $\mu = \sqrt{\zeta} = Q = 5$ GeV and $x=0.2, \, 0.3, \, 0.4$ (from left to right), normalized at $|\boldsymbol{k}_\perp|=0$. They are compared with the TMD PDF obtained in the flavor-independent MAPTMDPion25 FI fit. All the uncertainty bands correspond to the central 68\% C.L. The main outcome from the flavor-dependent analysis is the possibility to have an evident difference between the up and the down TMD PDFs, the former representing the sea-quark contribution and the latter describing the valence channel. The shapes of the $|\boldsymbol{k}_\perp|$ distribution are very different, the valence one having a wider tail (particularly, at large $x$) and the sea-quark one showing very big uncertainties at small $|\boldsymbol{k}_\perp|$. These details could not be exposed with the flavor-independent analysis, which produces just the single 
red band in Fig.~\ref{f:TMDPDF_pion_FD_norm}. 

\begin{figure}[h]
    \centering
    \includegraphics[width=0.99\linewidth]{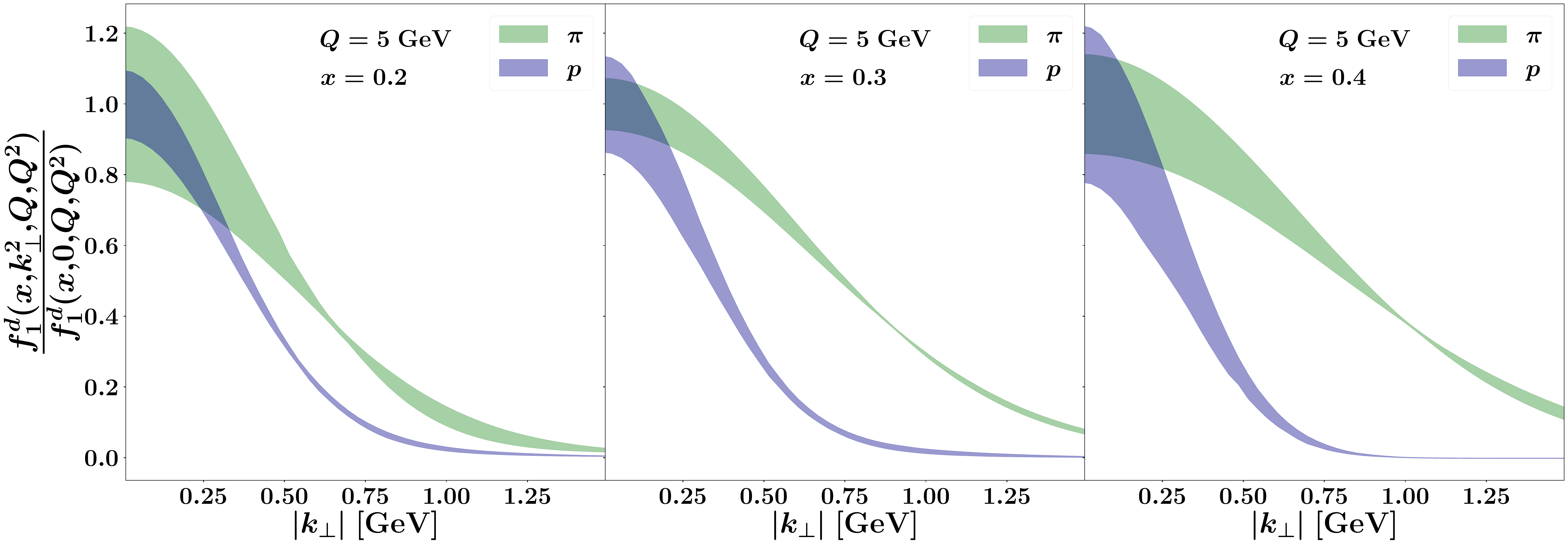}
    \caption{The $d$ quark TMD PDF as a function of $|\boldsymbol{k}_\perp|$ at $\mu = \sqrt{\zeta} = Q = 5$ GeV and $x=0.2, \, 0.3, \, 0.4$ (from left to right), normalized at 
    $\boldsymbol{k}_\perp|=0$. Blue (darker) band for the proton, green (lighter) band for $\pi^-$. All uncertainty bands correspond to the central 68\% C.L.}
    \label{f:TMDPDF_pion_proton_norm}
\end{figure}

In Fig.~\ref{f:TMDPDF_pion_proton_norm} we compare the TMD PDFs of the $d$ quark in $\pi^-$ (green/lighter band) and in the proton (blue/darker band), as functions of $|\boldsymbol{k}_\perp|$ at $\mu = \sqrt{\zeta} = Q = 5$ GeV and $x=0.2, \, 0.3, \, 0.4$ (from left to right), normalized at $|\boldsymbol{k}_\perp|=0$. It is evident from the plots that for all the explored $x$-values the TMD PDF in the pion has a larger $|\boldsymbol{k}_\perp|$-distribution than in the proton, the difference becoming sizable for increasing $x$. This outcome confirms the finding of Ref.~\cite{Schweitzer:2010tt,Barry:2023qqh}, where the pion TMD PDF turns out narrower than in the proton in Fourier-conjugate space. 

\begin{figure}[h]
    \centering
    \includegraphics[width=0.99\linewidth]{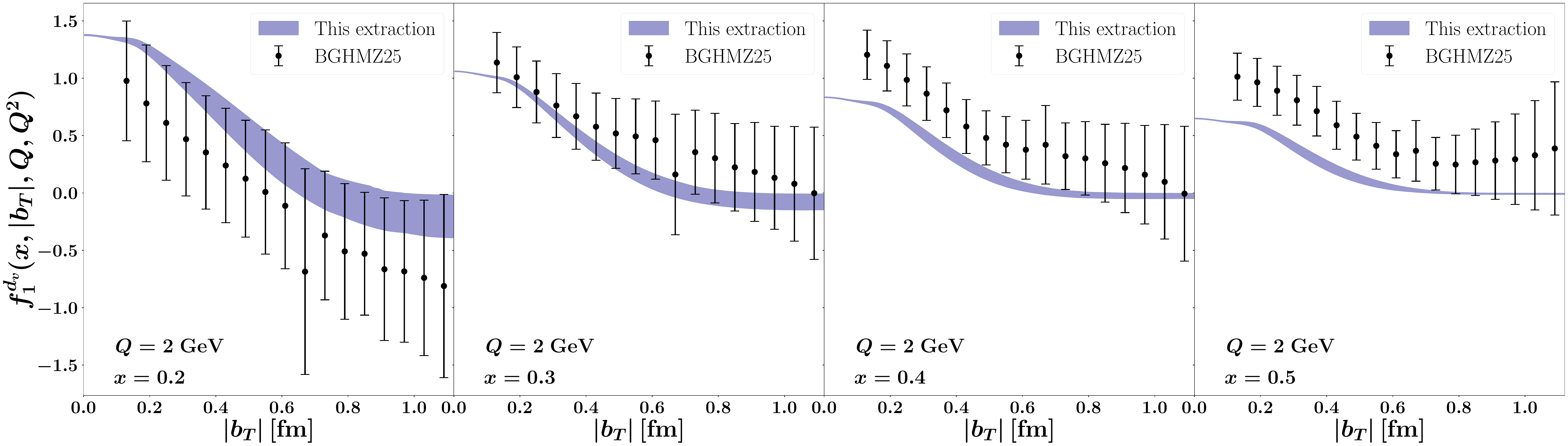}
    \caption{The valence $d_v = d - {\rm sea}$ quark TMD PDF in $\pi^-$ as a function of $|\boldsymbol{b}_T|$ at $\mu = \sqrt{\zeta} = Q = 2$ GeV and $x=0.2, \, 0.3, \, 0.4, \, 0.5$ (from left to right). Uncertainty band for the central 68\% C.L. from this work; black dots from lattice calculation of Ref.~\cite{Bollweg:2025iol} at $P^z = 3.04$ GeV.}
    \label{f:TMDPDF_lattice}
\end{figure}

Finally, in Fig.~\ref{f:TMDPDF_lattice} we compare our TMD PDF for the valence $d_v = d-\bar{d} \equiv d - {\rm sea}$ quark in $\pi^-$ with the same quantity computed on the lattice~\cite{Bollweg:2025iol} at $P^z = 3.04$ GeV, as function of $|\boldsymbol{b}_T|$ (Fourier-conjugated to $|\boldsymbol{k}_\perp|$) at $Q=2$ GeV and for $x=0.2, 0.3, 0.4, 0.5$ (from left to right). At $x=0.2$ (leftmost panel), lattice calculations are less reliable because systematic errors are very large (see also Fig.~23 of Ref.~\cite{Bollweg:2025iol}). The extracted TMD PDF is also less constrained because of a scarce coverage of data in this portion of phase space (see Fig.\ref{f:kin_coverage_pion}). At larger $x\gtrsim 0.3$, lattice calculations are more accurate, and the phenomenological extraction more constrained by data: both lattice error bars and TMD PDF uncertainty bands shrink. At the largest explored $x$, we observe a tendency to a slight disagreement, particularly at small $|\bT|$. We remark that for $|\bT| \lesssim 0.2$ fm the TMD PDF is basically driven by its perturbative component (Eq.~\eqref{e:pert_TMD}), which is  described at N$^3$LL accuracy. This trend deserves further studies. At $x=0.3$, the agreement is robust and encouraging for future comparisons with lattice calculations. 

\section{Conclusions}

In this work, we have performed for the first time the extraction of unpolarized quark TMDs in the pion with sensitivity of the  nonperturbative transverse momentum distribution of quarks to their flavor. We have analyzed all DY data involving pions and surviving kinematic cuts that satisfy TMD factorization, working at N$^3$LL$^-$ perturbative accuracy. We have performed the error analysis using the bootstrap method, including also the collinear PDF uncertainties propagated through Monte Carlo replicas. We reach a comparable or slightly better quality of the fit with respect to the flavor-independent analysis, namely $\chi^2_0 / N_{\text{cut}}=1.22$ vs. 1.28, where $\chi_0^2$ is referred to the best fit of the unfluctuated data. 

The limited set of DY data with pions, and their mild sensitivity to flavor, allow to separate only the $d=\bar{u}$ quark TMD from a cumulative sea-quark channel representing all other flavors. Nevertheless, the resulting TMDs show the possibility to have marked differences with respect to the flavor-independent analysis. The valence one has a wider tail at large transverse momenta, a feature persisting also in the comparison with the valence TMD in the proton for the same flavor, $x$ and hard scale $Q$, and it nicely compares with lattice calculations of the same quantity in position space and at intermediate $x$. This result could not be exposed by the simple flavor-independent analysis, and it represents the most important outcome of this work. The sea-quark TMD shows very large uncertainties at small transverse momenta because the corresponding fitting parameters are not tightly constrained by data, in particular by data from the E615 experiment which is plagued by large normalization errors. 

Upcoming COMPASS DY data will compensate the low coverage of experimental data in the $x_\pi \sim 0.2$ region of phase space. At present, the overall limited sensitivity of the current datasets to the nonperturbative component of the pion TMDs highlights the need for additional data to better constrain it and achieve a more detailed and insightful comparison between the partonic structures of the pion and the proton, the simplest bound states that can be formed in QCD.

\section{Acknowledgments}
This work is partially supported by the European Union ``Next
Generation EU'' program and the Italian Ministero dell'Universit\`a e Ricerca (MUR) through the PRIN2022 research grants 20229KEFAM (CUP H53D23000980006) and 
20225ZHA7W (CUP F53D23001320006).

\bibliography{biblio2}
\end{document}